# Towards Characterizing Markov Equivalence Classes for Directed Acyclic Graphs with Latent Variables

R. Ayesha Ali[1], Thomas S. Richardson[2], Peter Spirtes[3], Jiji Zhang[3]
[1]Dept. of Statistics & Applied Probability, National University of Singapore, Singapore
[2]Dept. of Statistics, University of Washington, Seattle, WA, USA
[3]Dept. of Philosophy, Carnegie Mellon University, Pittsburgh, PA, USA
staara@nus.edu.sg

## Abstract

It is well known that there may be many causal explanations that are consistent with a given set of data. Recent work has been done to represent the common aspects of these explanations into one representation. In this paper, we address what is less well known: how do the relationships common to every causal explanation among the observed variables of some DAG process change in the presence of latent variables? Ancestral graphs provide a class of graphs that can encode conditional independence relations that arise in DAG models with latent and selection variables. In this paper we present a set of orientation rules that construct the Markov equivalence class representative for ancestral graphs, given a member of the equivalence class. These rules are sound and complete. We also show that when the equivalence class includes a DAG, the equivalence class representative is the essential graph for the said DAG.

## 1 INTRODUCTION

Directed acyclic graph (DAG) models, represented by graphs consisting of vertices and directed ($\longrightarrow$) edges, encode the conditional independence relations holding among the variables (vertices) of some data generating process. Such models have been used in various forms such as path analyses in the social sciences, structural equation models in economics, and more recently as Bayesian networks in artificial intelligence. DAG models have many associated benefits, two main benefits being that associated with each DAG is $i$) a natural factorization of the joint density of the variables in the graph, and $ii$) a simple causal interpretation of the modelled process.

However, given a set of conditional independence relations, there are often many DAGs that can encode the same relations. All DAGs that encode the same set of conditional independence relations are *Markov equivalent*. Frydenberg (1990), Verma and Pearl (1991), Chickering (1995), Meek (1995) and Andersson et al. (1997) have characterized Markov equivalence classes for DAGs, and have presented algorithms for constructing an equivalence class representation given a member (DAG) of the equivalence class.

Following Andersson et al. (1997), we refer to the DAG equivalence class representative as the *essential graph*. For data generated by some DAG with no latent variables, one could correctly specify the associated essential graph. However, without knowing the underlying graph, one may worry that there were latent variables present, and that the learned essential graph is no longer valid. Ancestral graphs provide a class of graphs that can encode conditional independence relations that arise in DAG models with latent and selection variables. In this paper we present an algorithm for constructing a Markov equivalence class representative for ancestral graphs, given a single member of the class.

Section 2 provides relevant definitions and results on DAGs and Markov equivalence. Section 3 provides a unique representation of equivalence classes for maximal ancestral graphs. Analogous to Meek (1995), we provide in Section 4 a set of orientation rules that constructs the equivalence class representative for maximal ancestral graphs (see Definition 2.3). We prove that the orientation rules are sound and complete. We also show that whenever the equivalence class includes a DAG, the corresponding equivalence class representative is simply the DAG's corresponding essential graph.

## 2 BACKGROUND

A graphical Markov model is a pair $\langle V, E \rangle$ that represents an independence model where $V$ is a set of vertices (or variables); $E$ is a set of edges; and an *independence model* is a list of conditional independence statements such as "$A$ is independent of $B$ given $S$" for disjoint subsets, $\{A, B, S\}$, of $V$ and we write "$A \perp\!\!\!\perp B | S$".

### 2.1 VERTEX RELATIONS

We only consider graphs that have at most one edge between each pair of vertices. If there is an edge $\alpha \longrightarrow \beta$, or $\alpha \longleftrightarrow \beta$ then the *edge end at $\beta$ is an arrowhead*. Conversely, if there is an edge $\alpha \longrightarrow \beta$, or $\alpha \longrightarrow \beta$ then the *edge end at $\alpha$ is a tail*. We do not allow a vertex to be adjacent to itself. A *path*, $\pi$, is a sequence of distinct vertices that are adjacent.

If $\alpha$ and $\beta$ are vertices in a graph $\mathcal{G}$ such that $\alpha \longleftrightarrow \beta$, then $\alpha$ is a *spouse* of $\beta$ and vice versa. If $\alpha \longrightarrow \beta$ in $\mathcal{G}$, then $\alpha$ is a *parent* of $\beta$, and $\beta$ is a *child* of $\alpha$. If there is a directed path from $\alpha$ to $\beta$ (i.e. $\alpha \longrightarrow \longrightarrow \ldots \longrightarrow \beta$) or $\alpha = \beta$, then $\alpha$ is an *ancestor* of $\beta$, and $\beta$ is a *descendant* of $\alpha$.

A graph in which every edge is undirected forms an *undirected graph*. The skeleton of a graph $\mathcal{G}$ is the undirected graph formed by removing all arrowheads from $\mathcal{G}$.

### 2.2 MAXIMAL ANCESTRAL GRAPHS

#### 2.2.1 Directed Acyclic Graphs

A directed acyclic graph (DAG) is a graph such that all edges are directed ($\longrightarrow$), and there are no directed cycles. We say that a triple of vertices $\{x, y, z\}$ forms an *unshielded triple* if the pairs $(x, y)$ and $(y, z)$ are adjacent, but $x$ and $z$ are not adjacent. Otherwise, the triple is shielded and forms a *triangle*. For DAGs, a non-endpoint vertex $v$ on a path is said to be a *collider* if two arrowheads meet at $v$ (i.e. $\longrightarrow v \longleftarrow$); all other non-endpoint vertices on a path are *non-colliders*, (i.e. $\longrightarrow v \longrightarrow$, $\longleftarrow v \longrightarrow$). The independence relations entailed by a DAG can be determined through *d-separation*.

**Definition 2.1** *In a directed acyclic graph, a path $\pi$ between $\alpha$ and $\beta$ is said to be d-connecting given $Z$ if the following hold:*

*(i) No non-collider on $\pi$ is in $Z$;*
*(ii) Every collider on $\pi$ is an ancestor of a vertex in $Z$.*

*Two vertices $\alpha$ and $\beta$ are said to be d-separated given $Z$ if there is no path d-connecting $\alpha$ and $\beta$ given $Z$.*

In particular, if vertices $\alpha$ and $\beta$ are d-separated given $Z$, then $\alpha$ is independent of $\beta$ conditional on $Z$. However, for processes in which (a) some variables in the DAG are not observed ('latent'); or (b) other variables, specifying the specific subpopulation from which our data is sampled, are conditioned upon ('selection variables'); the independence model obtained by conditioning on the selection variables and marginalizing over the latent variables cannot be represented by a DAG, in general, even though the full underlying model can.

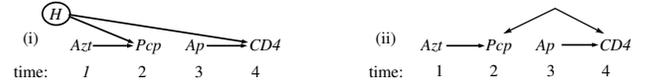

**Figure 1:** (i) A DAG with a latent variable $H$. (ii) The ancestral graph resulting from marginalizing over $H$ adds a bi-directed edge between $Pcp$ and $CD4$.

The DAG in Figure 1(i) entails the relation that $Azt \perp\!\!\!\perp CD4$; but $Azt \not\!\perp\!\!\!\perp CD4 | Pcp$. Similarly, $Ap \perp\!\!\!\perp Pcp$; but $Ap \not\!\perp\!\!\!\perp Pcp | CD4$. There is no DAG over the variables $\{Azt, Pcp, Ap, CD4\}$ that can simultaneously encode these relations.

However, ancestral graphs enable one to focus on the independence structure over the observed variables that results from the presence of latent variables without explicitly including latent variables in the model. Permitting bi-directed ($\longleftrightarrow$) edges in the graph allows one to graphically represent the existence of an unobserved common cause of observed variables. For Figure 1(i) this corresponds to removing $H$ from the graph and adding a bi-directed edge between $Pcp$ and $CD4$. Undirected edges ($\longrightarrow$) are also introduced to represent other unobserved (selection) variables that have been conditioned on. See Richardson and Spirtes (2003) for a detailed discussion on the interpretation of edges in an ancestral graph.

#### 2.2.2 Ancestral Graphs

**Definition 2.2** *A graph, which may contain undirected ($\longrightarrow$), directed ($\longrightarrow$) and bi-directed edges ($\longleftrightarrow$) is ancestral if:*

*(a) there are no directed cycles;*
*(b) whenever an edge $x \longleftrightarrow y$ is in the graph, then $x$ is not an ancestor of $y$, (and vice versa);*
*(c) if there is an undirected edge $x \longrightarrow y$ then $x$ and $y$ have no spouses or parents.*

The term 'ancestral' is motivated by conditions (a) and (b), which state that if $x$ and $y$ are joined by an edge and there is an arrowhead at $x$, then $x$ is *not* an ancestor of $y$. Condition (c) ensures that undirected edges never meet arrowheads ($\longrightarrow\gamma\!\relbar\!\relbar$, $\leftarrow\!\relbar\!\relbar\gamma\!\relbar\!\relbar$) in an ancestral graph. DAGs and undirected graphs form subclasses of ancestral graphs. McDonald (2002) considers a similar class of graphs, though without undirected edges, which he refers to as having ordered orthogonal errors (OOE).

Given an ancestral graph $\mathcal{G}$ with vertex set $V$, for arbitrary disjoint sets $S, L$ (both possibly empty) Richardson and Spirtes (2002) defined a graphical transformation such that the independence model corresponding to the transformed graph will be the independence model obtained by marginalizing over $L$ and conditioning on $S$ in the independence model of the original graph. Though this transformation is defined for any ancestral graph $\mathcal{G}$, the primary motivation is the case in which $\mathcal{G}$ is an underlying (causal) DAG that is partially observed. For ancestral graphs, a natural extension of the notions of 'collider' and 'non-collider' allows for the presence of undirected and bi-directed edges, i.e. ($\longrightarrow v \leftarrow\!\relbar\!\relbar$, $\leftarrow\!\relbar v \leftarrow\!\relbar\!\relbar$, $\leftarrow\!\relbar v \leftarrow\!\relbar\!\relbar$, $\longrightarrow v \leftarrow\!\relbar\!\relbar$); and ($\relbar\!\relbar v \relbar\!\relbar$, $\relbar\!\relbar v \longrightarrow$, $\longrightarrow v \longrightarrow$, $\leftarrow\!\relbar v \longrightarrow$) respectively.

Hence a natural extension of Pearl's d-separation criterion may be applied to ancestral graphs. In particular, *m-connection* and *m-separation* for ancestral graphs read like Definition 2.1 where 'collider' and 'non-collider' are as defined in the previous paragraph. Further, if $x$ is m-separated from $y$ given $Z$ in an ancestral graph, then we write $x \perp\!\!\!\perp_m y | Z$. Since m-separation characterizes the independence model entailed by an ancestral graph, tests of m-separation can be used to determine when graphs are Markov equivalent to each other (see Section 2.3).

Independence models described by DAGs satisfy pairwise Markov properties such that every missing edge corresponds to a conditional independence relation. In general, this property does not apply to ancestral graphs. For example, there is no set which m-separates $\gamma$ and $\delta$ in the graph in Figure 2(a), which motivates the following definition:

**Definition 2.3** *An ancestral graph $\mathcal{G}$ is said to be "maximal" if, for every pair of non-adjacent vertices $\alpha, \beta$ there exists a set $Z$, ($\{\alpha, \beta\} \notin Z$), such that $\alpha$ and $\beta$ are m-separated conditional on $Z$.*

These graphs are termed *maximal* in the sense that no additional edge may be added to the graph without changing the associated independence model. It has been shown in Richardson and Spirtes (2002) that if an ancestral graph is not maximal, then there exists at least one pair of non-adjacent vertices $\{\alpha, \beta\}$, for which there is an "inducing path" between $\alpha$ and $\beta$ where:

**Definition 2.4** *An inducing path $\pi$ is a path in a graph such that every non-endpoint vertex is a collider on the path, and an ancestor of at least one endpoint. If $\pi$ is the shortest inducing path, then $\pi$ is minimal.*

By definition, inducing paths always consist of a single edge in DAGs and in undirected graphs; hence, such graphs are always maximal. By adding a bi-directed edge between $\gamma$ and $\delta$, the non-maximal graph in Figure 2(a) can be made maximal, as shown in Figure 2(b). In the remainder of this paper, we focus on maximal ancestral graphs since every non-maximal ancestral graph can uniquely be transformed to a Markov equivalent maximal ancestral graph, by appropriately adding bi-directed edges.

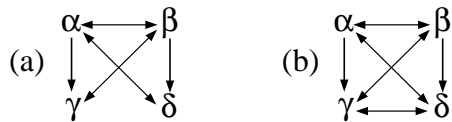

**Figure 2:** (a) The path $\{\gamma, \beta, \alpha, \delta\}$ is an example of an inducing path in an ancestral graph. (b) A maximal ancestral graph Markov equivalent to (a).

### 2.3 MARKOV EQUIVALENCE

**Definition 2.5** *Two graphs $\mathcal{G}_1$ and $\mathcal{G}_2$ are said to be Markov equivalent if for all disjoint sets $A, B, Z$ (where $Z$ may be empty), $A$ and $B$ are m-separated given $Z$ in $\mathcal{G}_1$ if and only if $A$ and $B$ are m-separated given $Z$ in $\mathcal{G}_2$.*

We say that graphs $\mathcal{G}_1$ and $\mathcal{G}_2$ are Markov equivalent if they entail the same independence model. If $\mathcal{G}$ is a maximal ancestral graph then we define $[\mathcal{G}]$ to be the class of maximal ancestral graphs Markov equivalent to $\mathcal{G}$, i.e. a *Markov equivalence class*. The *skeleton* of a graph is an undirected graph with the same adjacencies. Verma and Pearl (1991) proved that:

**Theorem 2.1** *(DAG Equivalence) Directed acyclic graphs $\mathcal{D}_1$ and $\mathcal{D}_2$ are Markov equivalent if and only if $\mathcal{D}_1$ and $\mathcal{D}_2$ have the same skeleton and the same unshielded colliders.*

A key difference between DAGs and ancestral graphs is that having the same adjacencies and unshielded colliders, though necessary, are not sufficient for Markov equivalence of ancestral graphs.

Consider the graphs shown in Figure 3: $\mathcal{G}_1$ and $\mathcal{G}_3$ contain the same adjacencies and the same unshielded colliders, but these two graphs are not Markov equivalent to each other: In $\mathcal{G}_1$, $x \perp\!\!\!\perp_m y | q$; but in $\mathcal{G}_3$, $x \not\perp\!\!\!\perp_m y | q$. In fact in any graph Markov equivalent to $\mathcal{G}_1$, $\langle q, \beta, y \rangle$ forms a shielded collider. (There is only one such graph, $\mathcal{G}_2$, so $\{\mathcal{G}_1, \mathcal{G}_2\}$ forms a Markov equivalence class.) However, in general, it is clearly not necessary that two graphs share *all* of the same shielded colliders in order for them to be Markov equivalent.

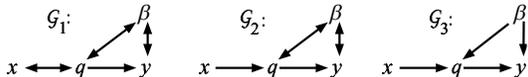

**Figure 3:** $\mathcal{G}_1$, $\mathcal{G}_2$, $\mathcal{G}_3$ have the same adjacencies and the same unshielded colliders, but $\mathcal{G}_1$ and $\mathcal{G}_3$ are not Markov equivalent. $\langle x, q, \beta, y \rangle$ forms a discriminating path for $\beta$ in every graph.

*Discriminating paths* are special paths that, if present in two Markov equivalent graphs, imply that certain shielded colliders (or non-colliders) will be present in both graphs:

**Definition 2.6** *A path* $\boldsymbol{\pi} = \langle x, q_1, q_2, \ldots, q_p, \beta, y \rangle$, *with $x$ not adjacent to $y$, is a* discriminating path for $\langle q_p, \beta, y \rangle$ *in an ancestral graph $\mathcal{G}$ if and only if for every vertex $q_i, 1 \leq i \leq p$ on $\boldsymbol{\pi}$, (i.e. excluding $x, y$, and $\beta$):*

(i) $q_i$ *is a collider on* $\boldsymbol{\pi}$; *and*
(ii) $q_i \longrightarrow y$, *hence forming a non-collider along the path* $\langle x, q_1, \ldots, q_i, y \rangle$.

Given a set $Z$, if $Z$ does not contain all $q_i, 1 \leq i \leq p$, then the path $\boldsymbol{\pi} = \langle x, q_1, \ldots, q_j, y \rangle$ is m-connecting where $q_j \notin Z$ and $q_i \in Z$ for all $i < j$. If $Z$ contains $\{q_1, \ldots, q_p\}$ and $\beta$ is a collider on the path $\boldsymbol{\pi}$ in the graph $\mathcal{G}$, then $\beta \notin Z$ if $Z$ m-separates $x$ and $y$. Consequently, in any graph Markov equivalent to $\mathcal{G}$ containing the discriminating path $\boldsymbol{\pi}$, $\beta$ is also a collider on $\boldsymbol{\pi}$. Conversely, if $\beta$ is a non-collider on the path $\boldsymbol{\pi}$ then $\beta$ is a member of any set that m-separates $x$ and $y$, and $\beta$ is a non-collider on $\boldsymbol{\pi}$ in any graph Markov equivalent to $\mathcal{G}$ containing $\boldsymbol{\pi}$. In other words, $\langle q_p, \beta, y \rangle$ is "discriminated" to be either a collider or a non-collider on the path $\boldsymbol{\pi}$ in any graph Markov equivalent to $\mathcal{G}$ in which $\boldsymbol{\pi}$ forms a discriminating path, even though it is shielded. The paths $\langle x, q, \beta, y \rangle$ in $\mathcal{G}_1$ and $\mathcal{G}_2$ from Figure 3 are examples of discriminating paths for $\beta$.

It is clear that discriminating paths, when present in both graphs, lead directly to necessary conditions for Markov equivalence. However, a discriminating path for a given triple may not be present in all graphs within a Markov equivalence class. We avoid this problem by identifying, via a recursive definition, a subclass of discriminating paths (those 'with order') which are always present. In particular, define a hierarchy of triples as follows:

**Definition 2.7** *Let $\mathfrak{O}_i$ $(i \geq 0)$ be the set of* triples of order $i$, *defined recursively as follows:*

Order 0: *A triple $\langle \alpha, \beta, \gamma \rangle \in \mathfrak{O}_0$ if $\alpha$ and $\gamma$ are not adjacent in $\mathcal{G}$.*

Order $i+1$: *A triple $\langle \alpha, \beta, \gamma \rangle \in \mathfrak{O}_{i+1}$ if*

*(1) $\langle \alpha, \beta, \gamma \rangle \notin \mathfrak{O}_j$, for some $j < i+1$ and*
*(2) there exists a discriminating path $\boldsymbol{\pi} = \langle x, q_1, \ldots, q_p = \alpha, \beta, \gamma \rangle$ for $\beta$ in $\mathcal{G}$, and each of the colliders on the path:*
$\langle x, q_1, q_2 \rangle, \ldots \langle q_{p-1}, q_p, \beta \rangle \in \bigcup_{j \leq i} \mathfrak{O}_j$.

*If $\langle \alpha, \beta, \gamma \rangle \in \mathfrak{O}_i$ then the triple is said to have order $i$. A discriminating path is said to have order $i$ if every triple on the path has order at most $i$, and at least one triple has order $i$. If a triple has order $i$ for some $i$, then we will say that the triple* has order, *likewise for discriminating paths.*

In each graph in Figure 3, the triple $\langle x, q, \beta \rangle$ has order 0, while $\langle q, \beta, y \rangle$ has order 1. It is important to note that not every triple in a graph will have order. For example, no triple in Figure 2(b) has order. Ali et al. (2004) proved the following Markov equivalence result for maximal ancestral graphs.

**Theorem 2.2** *Maximal ancestral graphs $\mathcal{G}_1$ and $\mathcal{G}_2$ are Markov equivalent if and only if $\mathcal{G}_1$ and $\mathcal{G}_2$ have the same adjacencies and colliders with order.*

## 3 JOINED GRAPHS

Andersson et al. (1997) constructed the essential graph, a Markov equivalence class representation for DAGs, by retaining all edges common to every member of the equivalence class. In other words, the essential graph associated with DAG $\mathcal{D}$ is a partially directed graph with the same skeleton as $\mathcal{D}$ such that an edge along a particular path is oriented if and only if the edge has the same orientation in every DAG in the equivalence class. Also, the essential graph for $[\mathcal{D}]$ is a chain graph (see Andersson et al. (1997), Frydenberg (1990)) and is Markov equivalent to every DAG in the equivalence class.

Note that the arrowheads in the essential graph form a subset of the arrowheads that were present in $\mathcal{D}$. Suppose that one could correctly specify the essential graph associated with some DAG. One may subsequently worry that if we allow for the presence of latent variables, then the essential graph may no longer entail all and only the arrowheads present in the entire equivalence class. However, it turns out that even in the presence of latent variables, the essential graph remains correct (see Theorem 4.3).

The join operation identifies the features common to a set of Markov equivalent ancestral graphs and can be thought of as an AND operation on the "arrowheads" of the set of graphs being joined, and an OR operation on the "tails" of these graphs. Following Andersson et al. (1997), Ali and Richardson (2002) made the following definition to join two maximal ancestral graphs:

**Definition 3.1** *If $\mathcal{G}_1$ and $\mathcal{G}_2$ are two graphs with the same adjacencies then define $\mathcal{G}_1 \vee \mathcal{G}_2$ to be a graph with the same adjacencies such that, on an edge between $\alpha$ and $\beta$, there is an arrowhead at $\beta$ in $\mathcal{G}_1 \vee \mathcal{G}_2$ if and only if there is an arrowhead at $\beta$ in $\mathcal{G}_1$ and in $\mathcal{G}_2$.*

Note that if two maximal ancestral graphs are Markov equivalent, then by Theorem 2.2 they will have the same adjacencies. We also define:

$$\sup[\mathcal{G}] = \bigvee_{\mathcal{G}' \in [\mathcal{G}]} \mathcal{G}'.$$

Hence $\sup[\mathcal{G}]$ is a simple representation of Markov equivalence classes for maximal ancestral graphs. Ali and Richardson (2002) and Ali and Richardson (2004) defined Markov properties for joined graphs and proved that $\sup[\mathcal{G}]$ is in fact Markov equivalent to every ancestral graph in the equivalence class. In general, joined graphs are not ancestral.

### 3.1 INFERRING EDGE ENDS

We now define "invariance" and present results used to prove the results in Section 4.

**Definition 3.2** *If the edge end at $b$, on an edge $(a, b)$ in a maximal ancestral graph, is of the same type (tail, arrowhead) in every graph in the equivalence class, then the edge end at $b$ is "invariant". If the edge ends at $a$ and $b$ are invariant, then we say that the $(a, b)$ edge is invariant.*

Partial characterizations of Markov equivalence classes for ancestral graphs have been obtained using partial ancestral graphs (PAGs) by Richardson and Spirtes (2003) and Spirtes et al. (1993). Unlike joined graphs, PAGs track both arrowheads and tails that are invariant in an equivalence class. Spirtes et al. (1999) also developed the Fast Causal Inference (FCI) algorithm to construct a PAG to represent a set of features common to every graph in $[\mathcal{G}]$, but this algorithm is not complete.

We use the following notation for edge ends of an edge $(a, b)$ in any graph $\mathcal{G}$:

1. "$a \relbar\joinrel\mathrel? b$" denotes that there is a tail at $a$, and either a tail or arrowhead at $b$ in $\mathcal{G}$.

2. "$a \mathrel{\prec\joinrel\mathrel?} b$" denotes that there is an arrowhead at $a$, and either a tail or arrowhead at $b$ in $\mathcal{G}$.
3. "$a \mathrel? \joinrel\mathrel? b$" denotes that there is either a tail or arrowhead at either edge end in $\mathcal{G}$.

Note that the above notation is merely a shorthand since we only consider graphs with edges that are directed, bi-directed or undirected. The following results infer arrowheads or tails in ancestral graphs.

**Lemma 3.1** *Let $\mathcal{G}$ be an ancestral graph containing vertices $\{a, b, c\}$. If, in $\mathcal{G}$, $a \mathrel? \joinrel\rightarrow b \longrightarrow c \mathrel? \joinrel\mathrel? a$, or $a \longrightarrow b \mathrel? \joinrel\rightarrow c \mathrel? \joinrel\mathrel? a$ then $c \mathrel{\prec\joinrel\mathrel?} a$.*

**Proof**: If $c \relbar\joinrel\mathrel? a$ in $\mathcal{G}$, then $a \mathrel? \joinrel\rightarrow b \longrightarrow c \relbar\joinrel\mathrel? a$ or $a \longrightarrow b \mathrel? \joinrel\rightarrow c \relbar\joinrel\mathrel? a$ violates the ancestral property. □

**Lemma 3.2** *If $\pi = \langle x, q_1, \ldots, q_p, b, y \rangle$ forms a discriminating path for $b$ given $Z$ in an ancestral graph $\mathcal{G}$, and $\langle q_p, b, y \rangle$ is a non-collider, then $b \longrightarrow y$ in $\mathcal{G}$.*

**Proof**: Since $\pi$ forms a discriminating path for $b$ given $Z$, $b \mathrel? \joinrel\rightarrow q_p \longrightarrow y \mathrel? \joinrel\mathrel? b$ is in $\mathcal{G}$. By Lemma 3.1, $b \mathrel? \joinrel\rightarrow y$; and since $\langle q_p, b, y \rangle$ is a non-collider, $b \leftarrow\joinrel\rightarrow y$ is ruled out: $y \mathrel{\prec\joinrel\rightarrow} b \longrightarrow q_p \longrightarrow y$ violates the ancestral property. Hence $b \longrightarrow y$ in $\mathcal{G}$. □

**Corollary 3.1** *If $\pi = \langle x, q_1, \ldots, q_p, b, y \rangle$ forms a discriminating path with order for $b$ in a maximal ancestral graph $\mathcal{G}$, and $\langle q_p, b, y \rangle$ is a non-collider in $\mathcal{G}$, then $b \longrightarrow y$ in every maximal ancestral graph Markov equivalent to $\mathcal{G}$ and the $(b, y)$ edge is invariant.*

**Proof**: By Markov equivalence, the path analogous to $\pi$ forms a discriminating path for $b$ and $\langle q_p, b, y \rangle$ is a non-collider in every graph in the equivalence class. By Lemma 3.2 $b \longrightarrow y$ in every maximal ancestral graph Markov equivalent to $\mathcal{G}$. □

**Lemma 3.3** *If $\langle u, a, w \rangle$ is a collider in the ancestral graph $\mathcal{G}$, $\langle u, b, w \rangle$ is a non-collider in $\mathcal{G}$, and $a$ is adjacent to $b$, then $a \mathrel{\prec\joinrel\mathrel?} b$ in $\mathcal{G}$.*

**Proof**: Since $\langle u, b, w \rangle$ is a non-collider, either $b \longrightarrow u \mathrel? \joinrel\rightarrow a \mathrel? \joinrel\mathrel? b$ or $b \longrightarrow w \mathrel? \joinrel\rightarrow a \mathrel? \joinrel\mathrel? b$ in $\mathcal{G}$ and by Lemma 3.1 $a \mathrel{\prec\joinrel\mathrel?} b$. □

## 4 CONSTRUCTING $\sup[\mathcal{G}]$

Meek (1995) presented a set of orientation rules that could be applied to a DAG to construct its associated essential graph. We now define a set of orientation rules that can be applied to a maximal ancestral graph $\mathcal{G}$ to construct $\sup[\mathcal{G}]$. See Ali et al. (2005) for full proofs of all results presented in this section.

**Orientation Rules for $[\mathcal{G}]$**

(S1) Let $\mathcal{H}$ be the skeleton of $\mathcal{G}$.

(S2) For all triples $x, y, z$, if $\langle x, y, z \rangle$ forms an unshielded collider in $\mathcal{G}$, then orient $x \text{?}{\rightarrow} y {\leftarrow}\text{?} z$ in $\mathcal{H}$.

(S3) If $\langle x, q_1, \ldots q_p, b, y \rangle$ forms a discriminating path for $b$ in $\mathcal{H}$, and $\langle q_p, b, y \rangle$ forms a collider in $\mathcal{G}$ then then $q_p\text{?}{\rightarrow} b {\leftarrow}\text{?} y$ in $\mathcal{H}$.

(S4) If $\langle u, a, v \rangle$ forms an unshielded collider in $\mathcal{H}$, and $\langle u, b, v \rangle$ forms an unshielded non-collider in $\mathcal{H}$, and $a$ and $b$ are adjacent then add an arrowhead at $a$ to the $(a, b)$ edge in $\mathcal{H}$: $a {\leftarrow}\text{?} b$.

(S5) If *either* of the following hold:

  (S5i) $\langle a, b, c \rangle$ forms an unshielded non-collider in $\mathcal{G}$, and $a\text{?}{\rightarrow} b$ in $\mathcal{H}$; or

  (S5ii) $\langle x, q_1, \ldots q_p \equiv a, b, c \rangle$ forms a discriminating path for $b$ in $\mathcal{H}$ and $\langle a, b, c \rangle$ forms a non-collider in $\mathcal{G}$;

  then perform the following orientations in $\mathcal{H}$:

  (S5a) Orient $b {\longrightarrow} c$.
  (S5b) For every vertex $z$ adjacent to $b$ and $c$, if $b {\leftarrow}\text{?} z$ in $\mathcal{H}$, then orient $z \text{?}{\rightarrow} c$.
  (S5c) For every vertex $z$ adjacent to $b$ and $c$, if $c \text{?}{\rightarrow} z$ in $\mathcal{H}$, then orient $z {\leftarrow}\text{?} b$.

(S6) Iterate steps (S3) to (S5) until no further arrowheads are added.

**Theorem 4.1** *The orientation rules are sound.*

The proof proceeds by showing that all arrowheads introduced by the orientation rules are invariant in $[\mathcal{G}]$. By definition, the graph resulting from joining the entire equivalence class will also contain these arrowheads. Hence $\sup[\mathcal{G}]$ contains all the arrowheads introduced by the orientation rules and the procedure is sound.

The following concept is central to showing that the orientation rules are complete (see Theorem 4.2).

**Definition 4.1** *(Balanced) A triangle with vertex set $\{x, y, z\}$ is said to be balanced at $x$ if one of the following holds: (i) $y \text{?}{\rightarrow} x {\leftarrow}\text{?} z \text{?}{\text{---}}\text{?} y$; (ii) $y \text{?}{\text{---}} x {\text{---}}\text{?} z \text{?}{\text{---}}\text{?} y$; or (iii) $y \text{?}{\rightarrow} x {\text{---}}\text{?} z {\leftarrow}\text{?} y$.*

In summary, the triangle is balanced at $x$ if the edge ends at $x$ are of the same type (arrowhead or tail); if the edge ends differ, then the triangle is balanced if (*iii*) holds. If every vertex in a triangle is balanced, then the triangle is balanced. A graph containing directed, undirected and bi-directed edges will be said to be *balanced* if every triangle in the graph is balanced. It can easily be verified that ancestral graphs and DAGs are balanced.

**Lemma 4.1** *The graph $\mathcal{H}$ produced by the orientation rules is balanced.*

Suppose that $\mathcal{H}$ is not balanced. Then there is some triangle $\langle x, y, z \rangle$ in $\mathcal{H}$ in which $x \text{?}{\rightarrow} y {\text{---}}\text{?} z {\text{---}}\text{?} x$. The proof begins by considering the *first* arrowhead $x \text{?}{\rightarrow} y$ introduced by the orientation rules into a triangle $\langle x, y, z \rangle$ which *remains* unbalanced after the procedure has completed. Ali et al. (2005) show that each step of the procedure either could not have introduced an arrowhead into a triangle which remained unbalanced, *or* that the supposition that it did implies that there was already an arrowhead that had been introduced earlier into a triangle which remained unbalanced, which is also a contradiction.

Lemma 4.1 is powerful because it characterizes the types of configurations that arise as a result of the orientation rules. For instance, a direct corollary of Lemma 4.1 is that the endpoints of an undirected edge in $\mathcal{H}$, that was not undirected in the maximal ancestral graph $\mathcal{G}$ that gave rise to $\mathcal{H}$, share the same parents and spouses.

**Corollary 4.1** *Let $\mathcal{H}$ be the graph produced by the orientation rules. If in $\mathcal{H}$ either i) $a {\longrightarrow} b {\text{---}} c$, or ii) $a {\leftrightarrow} b {\text{---}}$, then $a {\longrightarrow} c$ or $a {\leftrightarrow} c$ respectively.*

**Proof**: $a$ is adjacent to $c$ else edge $b {\text{---}} c$ contradicts Step (S5a) of the orientation rules. The triangle is balanced at $b$ hence $a \text{?}{\rightarrow} c$ in $\mathcal{H}$. Similarly, the triangle is balanced at $b$ hence $a {\leftrightarrow} c$ if and only if $a {\leftrightarrow} b$. □

The following two corollaries will be instrumental in proving Theorem 4.2.

**Corollary 4.2** *Let $\mathcal{H}$ be the graph produced by the orientation rules. Suppose $\mathcal{H}^*$, the induced subgraph obtained by removing the undirected edges from $\mathcal{H}$, forms an ancestral graph. Then $\mathcal{H}^*$ is maximal.*

**Corollary 4.3** *Let $\mathcal{H}$ be the graph produced by the orientation rules. Then no replacement of the undirected edges in $\mathcal{H}$ by directed edges will give rise to an inducing path with non-adjacent endpoints, that includes an edge that was oriented by the orientation rules.*

Another key concept required to prove that the orientation rules are complete involves defining an *order* on the variables in a graph. A graph is *chordal* if and only if every cycle over four or more vertices has an edge between two non-adjacent vertices, i.e. has a chord. A *partial* order ($\preceq$) for a graph induces an orientation such that if $x \preceq y$, then there is no directed path from

$y$ to $x$ ($y$ is *not* an ancestor of $x$). For an undirected graph $\mathcal{U}$, a *total* order induces an orientation such that for $\{x,y\}$ adjacent, $x \longrightarrow y$ if and only if $x \preceq y$. Let $\mathcal{U}_\alpha$ be the induced directed graph obtained by a total order $\alpha$. Then we say that $\mathcal{U}$ has a *consistent* order $\alpha$ if and only if $\mathcal{U}_\alpha$ has no unshielded colliders. We make use of Meek's (1995) results for undirected graphs.

**Lemma 4.2** *For undirected graphs, only chordal graphs have consistent orderings.*

**Lemma 4.3** *(Orienting chordal graphs) Let $\mathcal{U}$ be an undirected chordal graph. For all pairs of adjacent vertices $x$ and $y$ in $\mathcal{U}$ there exist total orderings $\alpha$ and $\gamma$ which are consistent with respect to $\mathcal{U}$ and such that $x \longrightarrow y$ is in $\mathcal{U}_\alpha$ and $y \longrightarrow x$ is in $\mathcal{U}_\gamma$.*

**Theorem 4.2** *The orientation rules are arrowhead complete.*

**Proof**: Let $\mathcal{H}$ be the graph resulting from performing the orientation rules on maximal ancestral graph $\mathcal{G}$. There are four steps to the proof.

I. *Removing the undirected edges in $\mathcal{H}$ leaves a disjoint union of maximal ancestral graphs.* Let $\mathcal{H}^*$ be the graph resulting from removing the undirected edges in $\mathcal{H}$. Suppose for a contradiction that $\mathcal{H}^*$ contains a non-ancestral configuration. By construction, there are no undirected edges in $\mathcal{H}^*$. Hence configurations such as $a\;?\!\!\longrightarrow\!b\!\text{---}\!c$ or $a\;?\!\!\longrightarrow\!b\!\text{---}\!c\!\text{---}\!d\!\longrightarrow\!a$ do not occur in $\mathcal{H}^*$. Then $\mathcal{H}^*$ contains a partially directed k-cycle of the form $q_1\;?\!\!\longrightarrow\!q_2\!\longrightarrow\cdots\longrightarrow q_k\longrightarrow q_1$.

Consider $k = 3$. By Lemma 4.1, there are no partially directed cycles involving only three vertices in $\mathcal{H}^*$. By induction, $\mathcal{H}^*$ contains no partially directed k-cycles: $\mathcal{G}$ includes at least one collider in a k-cycle, and hence at least one triple in the k-cycle is shielded in $\mathcal{G}$. Suppose triple $\{q_{i-1}, q_i, q_{i+1}\}$ is shielded, $1 < i < k$. By Lemma 4.1, $q_{i-1}\;?\!\!\longrightarrow\!q_{i+1}$ in $\mathcal{H}$, and thus in $\mathcal{H}^*$ too. But then either $\langle q_1, q_2, \ldots, q_{i-1}, q_{i+1}, \ldots, q_k, q_1\rangle$ or $\langle q_{i-1}, q_i, q_{i+1}\rangle$ forms a shorter cycle in $\mathcal{H}^*$, which is a contradiction.

Hence $\mathcal{H}^*$ is ancestral (with no undirected edges) and by Corollary 4.2, $\mathcal{H}^*$ is maximal.

II. *No replacement of the undirected edges in $\mathcal{H}$ by directed edges will give rise to a partially directed cycle, an unshielded collider, a collider with order, or an inducing path with non-adjacent endpoints, that includes an edge that was oriented by the orientation rules.* By Corollary 4.1, no orientation of the undirected edges in $\mathcal{H}$ will give rise to a partially directed cycle or an unshielded collider. By Corollary 4.3, no orientation of the undirected edges in $\mathcal{H}$ will give rise to an inducing path with non-adjacent endpoints.

Suppose for a contradiction that we can replace, by directed edges, the undirected edges in $\mathcal{H}$ such that triple $\{a,b,c\}$ is discriminated to be a collider by path $\boldsymbol{\pi} = \langle q_0, q_1, ..., q_p = a, b, c\rangle$. Call this graph $\mathcal{H}^*$. Then every non-endpoint of $\boldsymbol{\pi}$ is a collider oriented by the orientation rules, and $a\!\leftarrow\!\!\longrightarrow\!b\!\leftarrow\!\!\longrightarrow\!c$. Then there exists an $i$, $1 \leq i \leq p$ such that $q_i\text{---}c$ in $\mathcal{H}$. Since $q_0\;?\!\!\longrightarrow\!q_1\text{---}?c$ is unshielded, by (S5i)(S5a) the rules would orient $q_1\longrightarrow q_n$. Further, by (S5ii)(S5a) $q_i\longrightarrow q_n$ (since $\langle q_0, q_1, ..., q_{i-1}, q_i, c\rangle$ forms a discriminating path for a non-collider with $q_{i-1}\!\leftarrow\!\!\longrightarrow\!q_i\text{---}?c$) for $1 \leq i \leq p$, which is a contradiction.

III. *Let $\mathcal{U}$ be the undirected graph obtained by removing from $\mathcal{H}$ the directed and bi-directed edges, as well as all undirected edges that have no parents or spouses. Then $\mathcal{U}$ is a disjoint union of chordal undirected graphs.* Suppose for a contradiction that $\mathcal{U}$ is not decomposable. Then all total orderings of $\mathcal{U}$ lead to an unshielded collider. By II., we know that no replacement of the undirected edges in $\mathcal{H}$ by directed edges gives rise to a collider with order or an inducing path with non-adjacent endpoints involving the edges oriented by the orientation rules. Consider $\mathcal{U}'$, the subgraph of $\mathcal{G}$ corresponding to $\mathcal{U}$. It follows from Corollary 4.1 that $\mathcal{U}$ is an induced subgraph of $\mathcal{H}$, and hence $\mathcal{U}'$ is an induced subgraph of $\mathcal{G}$. Hence by rule (S2), $\mathcal{U}'$ does not contain any unshielded colliders.

Let $\mathcal{D}$ be the skeleton of $\mathcal{U}'$. Construct a total order for $\mathcal{D}$ as follows: remove all bi-directed edges from $\mathcal{U}'$, which leaves a DAG. Find a total ordering compatible with this DAG and orient, as directed edges, the edges in $\mathcal{D}$ according to this ordering. Note that every arrowhead in $\mathcal{D}$ is present in $\mathcal{G}$ on the corresponding directed or bi-directed edge in $\mathcal{U}'$. If $\mathcal{U}$ is not chordal, then $\mathcal{D}$ contains an unshielded collider. But this unshielded collider is also present in $\mathcal{U}'$, which is a contradiction.

IV. By Lemma 4.3 there are at least two such orderings for every $(x,y)$ edge in $\mathcal{U}$: one in which $x \longrightarrow y$ and another in which $x \longleftarrow y$. Hence, $\mathcal{H}$ is maximally oriented and therefore the orientation rules are arrowhead complete. $\square$

**Theorem 4.3** *Let $\mathcal{D}$ be a DAG containing only observed variables $\mathbf{O}$, and $\mathcal{G}$ be a maximal ancestral graph over the same observed variables. Further let $\mathcal{H}$ be the graph resulting from applying the orientation rules to $\mathcal{G}$, and $\mathcal{E}$ be the essential graph resulting from applying Meek's rules for DAGs to $\mathcal{D}$. If $\mathcal{D} \sim \mathcal{G}$ then $\mathcal{E} = \sup[\mathcal{G}]$.*

By Markov equivalence, $\mathcal{D}$ and $\mathcal{G}$ entail the same set of unshielded colliders. Further, since $\mathcal{D}$ contains no bi-directed edges: (*i*) all discriminating paths in $\mathcal{D}$ discriminate non-colliders; and (*ii*) any discriminating path with order has order of at most one. The proof of

Theorem 4.3 proceeds by showing that the operations performed on the skeleton of $\mathcal{G}$ are equivalent to the steps performed on the skeleton of $\mathcal{D}$.

## 5 CONCLUSIONS

Ancestral graphs are a class of graphs that can represent the independence relations holding among the observed variables of a DAG model with latent and selection variables. Unfortunately, as with DAG models, there often are a number of ancestral graphs that can encode the same independence model. Joined graphs, which can extract the arrowheads common to Markov equivalent graphs, allow one to associate a unique graph with each ancestral graph equivalence class. In this paper we have presented a set of orientation rules that constructs the joined graph for an entire equivalence class based on a single ancestral graph $\mathcal{G}$. Further, we have shown that if $[\mathcal{G}]$ contains a DAG $\mathcal{D}$, then $\sup[\mathcal{G}]$ equals the essential graph for $[\mathcal{D}]$.

The completeness proof suggests a way to construct a member of the equivalence class that contains the minimal number of arrowheads. Also, Drton and Richardson (2004) presented an algorithm for fitting ancestral graphs. Hence we have the framework for conducting an efficient equivalence class search across maximal ancestral graphs. The authors are currently working on this problem.

**Acknowledgments**

The authors are thankful to Michael Perlman, for posing the question of whether the essential graph for DAG $\mathcal{D}$ is equal to $\sup[\mathcal{G}]$ if $\mathcal{D} \in [\mathcal{G}]$, and to the William and Flora Hewlett Foundation and the Center for Advanced Studies in the Behavioral Sciences at Stanford University, where Thomas Richardson was a Fellow from 2003-2004. This research was supported by the National University of Singapore.

## References

Ali, R. A. and T. Richardson (2002). Markov equivalence classes for maximal ancestral graphs. In A. Darwiche and N. Friedman (Eds.), *Uncertainty in Artificial Intelligence: Proceedings of the* 18[th] *Conference*, San Francisco, pp. 1–8. Morgan Kaufmann.

Ali, R. A. and T. Richardson (2004). Searching across markov equivalence classes of maximal ancestral graphs. In *2004 Proceedings of the American Statistical Association, Statistical Computing Section [CD-ROM]*, Toronto, ON, pp. 2383–2391. American Statistical Association.

Ali, R. A., T. Richardson, and P. Spirtes (2004). Markov equivalence for ancestral graphs. Technical Report 466, Department of Statistics, University of Washington.

Ali, R. A., T. Richardson, P. Spirtes, and J. Zhang (2005). Orientation rules for constructing markov equivalence classes for maximal ancestral graphs. Technical Report 476, Department of Statistics, University of Washington.

Andersson, S. A., D. Madigan, and M. D. Perlman (1997). A characterization of Markov equivalence classes for acyclic digraphs. *Ann. Statist. 25*, 505–541.

Chickering, D. (1995). A transformational characterization of equivalent Bayesian networks. In P. Besnard and S. Hanks (Eds.), *Proceedings of the Eleventh Annual Conference on Uncertainty in Artificial Intelligence*, San Mateo, CA, pp. 87–98. Morgan Kaufmann.

Drton, M. and T. Richardson (2004). Iterative conditional fitting for gaussian ancestral graph models. In *Proceedings of the 20th Annual Conference on Uncertainty in Artificial Intelligence (UAI-04)*, Arlington, Virginia, pp. 130–137. AUAI Press.

Frydenberg, M. (1990). The chain graph Markov property. *Scand. J. Statist. 17*, 333–353.

McDonald, R. (2002). What can we learn from the path equations?: identifiability, constraints, equivalence. *Psychometrika 67*(2), 225–249.

Meek, C. (1995). Causal inference and causal explanation with background knowledge. In P. Besnard and S. Hanks (Eds.), *Uncertainty in Artificial Intelligence: Proceedings of the* 11[th] *Conference*, San Francisco, pp. 403–410. Morgan Kaufmann.

Richardson, T. and P. Spirtes (2003). Causal inference via ancestral graph models. In P. Green, N. Hjort, and S. Richardson (Eds.), *Highly Structured Stochastic Systems*. Oxford: Oxford University Press.

Richardson, T. S. and P. Spirtes (2002). Ancestral graph Markov models. *Annals of Statistics 30*(4), 962–1030.

Spirtes, P., C. Glymour, and R. Scheines (1993). *Causation, Prediction and Search*. Number 81 in Lecture Notes in Statistics. Springer-Verlag.

Spirtes, P., C. Meek, and T. S. Richardson (1999). Causal inference in the presence of latent variables and selection bias. In C. Glymour and G. F. Cooper (Eds.), *Computation, Causality and Discovery*, pp. 211–252. Cambridge, Mass.: MIT Press.

Verma, T. and J. Pearl (1991). Equivalence and synthesis of causal models. Technical Report R-150, Cognitive Systems Laboratory, UCLA.